\documentclass[aps,prb,twocolumn,superscriptaddress,notitlepage,floatfix]{revtex4-2}

\usepackage[T1]{fontenc}
\usepackage{times}

\usepackage{graphicx}
\usepackage{dcolumn}
\usepackage{bm}
\usepackage{amsmath,amssymb,amsthm}
\usepackage{braket}
\usepackage{multirow}
\usepackage{xcolor}
\usepackage[colorlinks,allcolors=blue]{hyperref}

\newcommand{\ii}{\mathrm{i}}
\newcommand{\dd}{\mathrm{d}}
\newcommand{\e}{\mathrm{e}}

\newcommand{\EF}{E_{\mathrm{F}}}
\newcommand{\tp}{t_{\perp}}

\begin{document}

\title{Parity-Anomalous Hall Response of Chiral Lifshitz Fermions}

\author{Yanmiao Han}
\affiliation{School of Physics, Beihang University, Beijing 102206, China}

\author{Rundong Zhao}
\affiliation{School of Physics, Beihang University, Beijing 102206, China}

\author{Yu-Hao Wan}
\email[Contact author:] {wanyh@stu.pku.edu.cn}
\affiliation{International Center for Quantum Materials and School of Physics, Peking University, Beijing 100871, China}


\begin{abstract}
The parity anomaly of a single massless Dirac fermion is manifested as a half-quantized Hall response once the required parity-breaking regularization is supplied. We study the corresponding response for two-dimensional chiral Lifshitz fermions. For a node with low-energy coupling $(q_x+\ii q_y)^N$, the Fermi-loop winding number $W=N$ gives an anomalous Hall response $\sigma_{xy}^{\mathrm{anom}}=(W/2)e^2/h$. We examine this mechanism in chiral multilayer Haldane models tuned to a valley-selective critical line, where a single gapless chiral Lifshitz fermion remains at the $K$ valley. For the AB, ABC, and ABCA stacks, the projected pseudospin textures give \(W=2,3,4\), while Kubo calculations yield \(\sigma_{xy}=(1,3/2,2)e^2/h\); integrated equilibrium boundary currents reproduce the same \(W/2\) sequence. Six-terminal simulations with dephasing further show that the Hall conductivity is quantized at $W/2$ while the longitudinal conductivity remains finite. These results identify gapless chiral Lifshitz fermions as a higher-order, winding-controlled generalization of the parity-anomalous Hall response, distinct from a quantum anomalous Hall insulating state even when $W/2$ is an integer.
\end{abstract}

\maketitle

\section{Introduction}

The parity anomaly is most sharply formulated for a single massless Dirac fermion in $(2+1)$ dimensions \cite{NiemiSemenoff1983PRL,Redlich1984PRD,Redlich1984PRL,Jackiw1984PRD}. At the classical level, the low-energy Dirac theory may preserve parity symmetry. After quantization, however, a regularization procedure is required, and maintaining gauge invariance generates a parity-odd Chern--Simons response with a half-quantized Hall coefficient \cite{NiemiSemenoff1983PRL,Redlich1984PRD,Redlich1984PRL,Semenoff1984PRL,FradkinDagottoBoyanovsky1986PRL}. In condensed matter, this  field-theoretic statement must be interpreted within a lattice band structure with a finite Brillouin zone. A strictly parity-symmetric single Dirac cone cannot be realized as an isolated two-dimensional lattice fermion because of fermion doubling \cite{NielsenNinomiya1981PLB,Karsten1981PLB,KarstenSmit1981NPB}. Haldane's honeycomb model provided the paradigmatic way around this obstruction: at the critical boundary between a Chern insulator and a trivial insulator, one valley is tuned to a massless point while the other valley remains gapped, so that the full lattice model supplies the necessary high-energy completion \cite{Haldane1988,Semenoff1984PRL,FradkinDagottoBoyanovsky1986PRL}. Recent studies have further clarified the origin of the half response in such lattice systems. In a parity-anomalous semimetal, the Hall conductance separates into an integer contribution from completely filled lattice bands and a Fermi-surface Berry-phase contribution from the gapless band \cite{Haldane2004PRL, Lee2009PRL,RosenbergGuoFranz2010PRB,ChuShiShen2011PRB,NomuraNagaosa2011PRL,Brune2011PRL,MulliganBurnell2013PRB,Konig2014PRB,Zhang2017NatCommun,Lapa2019PRB,Burkov2019PRB}. The half-quantized part is therefore tied to a parity-invariant massless Fermi surface, not to an isolated massive Dirac cone. This distinction is essential: once the full Brillouin zone and lattice regularization are properly included, a massive Dirac cone by itself does not carry a standalone half-quantized Hall conductance; the parity-anomalous response is instead a property of massless Dirac fermions whose low-energy parity symmetry is completed by parity-breaking high-energy lattice states \cite{Fu2022NPJ,Zou2022PRB,Zou2023PRB,ChenShen2024SCPMA,FuShen2025CommPhys,Shen2026PRB}.

This lattice viewpoint has guided a number of recent condensed-matter realizations and transport diagnostics of the parity anomaly. In semi-magnetic three-dimensional topological-insulator heterostructures, magnetic doping gaps one surface Dirac cone while leaving the opposite surface cone effectively massless, producing a half-quantized Hall response associated with the remaining parity-invariant cone \cite{Zhou2022PRL,Mogi2022NatPhys,Zou2023PRB}. Mirror-symmetric topological-insulator films offer a related route: mirror symmetry separates the system into two mirror sectors, each carrying an opposite half-quantized Hall response, whose difference gives a quantized mirror Hall effect \cite{Fu2024NatCommun}. In weak topological insulators, altermagnetism can {\color{black}gap} one of the doubled surface Dirac cones and leave a single massless cone, with the anomaly revealed through equilibrium current and Hall transport \cite{WanSun2025PRB}. More recently, analogous transport ideas have been extended to three-dimensional chiral topological phases, where an integer bulk winding number fixes the net chirality of boundary Dirac or Majorana cones and amplifies the corresponding surface anomaly coefficient \cite{WanHanZhaoSun2026Chiral}. These developments establish the parity anomaly as a measurable lattice phenomenon, visible in Hall conductance, boundary-current profiles, magnetic-field-driven crossovers, and multi-terminal transport \cite{Zou2022PRB,WangFuShen2024PRB,PhysRevB.109.195408,WangHuFuLiKongBaiShenXiao2026PRL}. Nevertheless, the low-energy building blocks considered so far are still massless Dirac fermions.

The conventional field-theoretic formulation of the parity anomaly is usually developed for relativistic Dirac fermions with linear dispersion and \(z=1\)\cite{DharMandalWadia2009,BakasLust2011,MontaniSchaposnik2012,AlexandreBristerHouston2012}. In broader quantum field theories, Lorentz invariance can be relaxed, giving rise to Lifshitz fermions characterized by anisotropic space-time scaling with $z \neq 1$. In condensed matter systems, such Lifshitz fermions emerge naturally as topological band crossings with nonlinear dispersions $E \propto q^N$ (where $z=N$) \cite{SunYaoFradkinKivelson2009,KoshinoMcCann2009,McCannKoshino2013,FangGilbertDaiBernevig2012}. A particularly intriguing class is the chiral Lifshitz fermions, where the higher-order dispersion is accompanied by a nontrivial pseudospin texture $(q_x+\mathrm{i}q_y)^N$ \cite{MinMacDonald2008}. Unlike a massless Dirac fermion, a chiral Lifshitz fermion carries an N-fold winding number around its Fermi loop. This fundamental geometric difference leaves open a compelling question: how does the parity anomaly manifest when the elementary building block is a massless chiral Lifshitz fermion rather than a massless Dirac cone? While the lattice parity anomaly and its transport signatures have been extensively studied for Dirac fermions, the corresponding anomalous transport driven by chiral Lifshitz fermions remains largely unexplored. Specifically, it remains to be explored how this higher-order winding translates into a macroscopic anomalous Hall response, and whether such an N-fold anomaly retains the characteristic metallic nature of the conventional Dirac case.

In this work, we propose a higher-order generalization of the parity anomaly driven by chiral Lifshitz fermions. We theoretically demonstrate that the anomalous Hall response of such a gapless cone is governed by the $N$-fold pseudospin winding of its parity-invariant Fermi loop, yielding an intrinsic metallic contribution $(N/2)e^2/h$. To verify this framework, we introduce multilayer Haldane models as a lattice realization. By tuning to a valley-selective critical line, we successfully isolate a single gapless higher-order cone. Through Kubo formula calculations, equilibrium boundary current evaluations, and dephasing-assisted multi-terminal transport simulations, we systematically confirm this winding-controlled anomaly. Crucially, our transport results reveal a quantized Hall plateau coexisting with a finite longitudinal conductivity, explicitly distinguishing this higher-order parity-anomalous metal from a fully gapped quantum anomalous Hall insulator.

The remainder of this paper is organized as follows. In Sec.~\ref{sec:higher-order-theory}, we formulate the low-energy theory of the chiral Lifshitz fermion, discuss the indispensable role of the parity-breaking high-energy lattice regularization, and analytically derive the winding-controlled anomalous Hall coefficient. Section~\ref{sec:haldane-realization} presents the microscopic lattice realization using multilayer Haldane stacks, detailing the numerical verification of the winding numbers and the multi-terminal transport signatures. Finally, Sec.~\ref{sec:summary-discussion} summarizes our findings and discusses their physical implications.

\section{Higher-Order Generalization of the Parity-Anomalous Hall Response}
\label{sec:higher-order-theory}
\begin{figure}[t]
    \includegraphics[width=\linewidth]{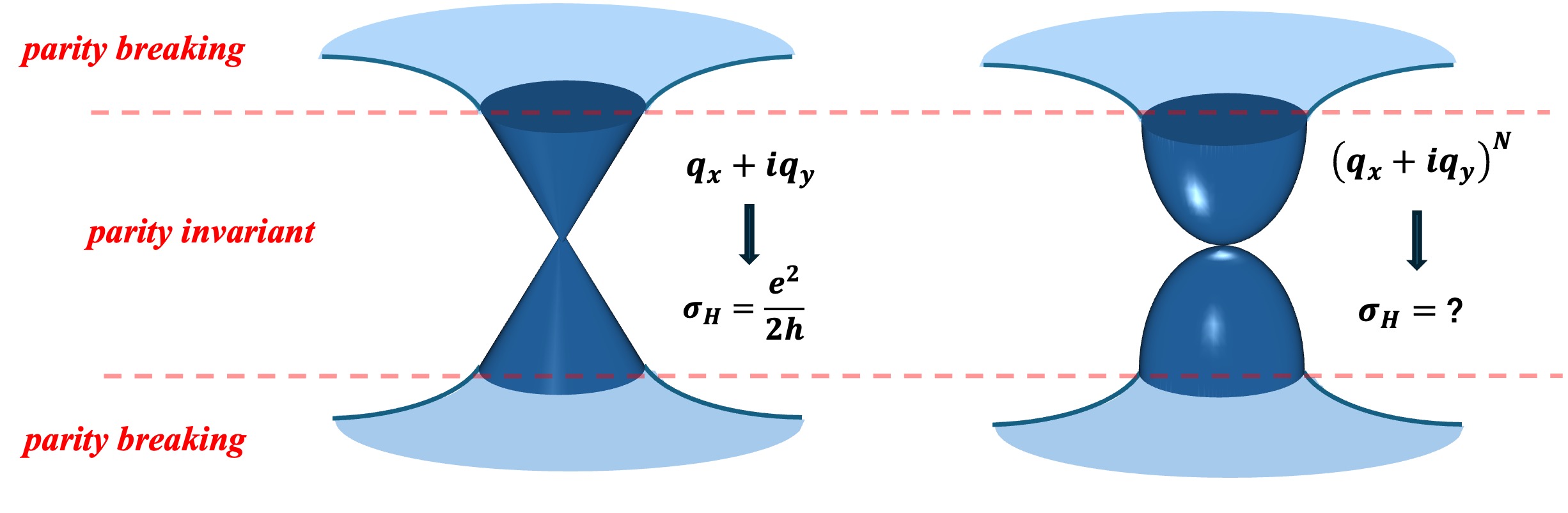}
    \caption{Hall response for a linear Dirac cone and a chiral Lifshitz fermion. The red dashed lines separate the parity-invariant low-energy window from the parity-breaking high-energy lattice sectors. For the linear cone with coupling $q_x+\ii q_y$, the anomalous response is $\sigma_H=e^2/2h$. For the chiral Lifshitz fermion with coupling $(q_x+\ii q_y)^N$, the response marked by the question mark is determined in Sec.~\ref{sec:higher-order-theory}.}
    \label{fig:fig1}
\end{figure}

Figure~\ref{fig:fig1} illustrates the conventional parity anomaly associated with a single massless Dirac cone \cite{NiemiSemenoff1983PRL,Redlich1984PRD,Redlich1984PRL,Semenoff1984PRL}. As shown in the left part of Fig.~\ref{fig:fig1}, the red dashed boundaries mark a low-energy window around the gapless point, where the effective Hamiltonian is well described by a massless Dirac cone and the Fermi surface preserves the local parity symmetry. The states outside this window belong to the high-energy part of the lattice spectrum and provide the lattice regularization of the low-energy Dirac theory. This distinction is essential in a crystalline system: the first Brillouin zone is finite and periodic, so a single Dirac cone cannot be extended over the whole lattice momentum space while preserving the same parity symmetry everywhere without conflicting with fermion doubling \cite{NielsenNinomiya1981PLB,Karsten1981PLB,KarstenSmit1981NPB,Haldane1988,Fu2024NatCommun}. A lattice completion must therefore contain high-energy regularization terms, such as Wilson-type lattice masses, which are negligible near the Dirac point but become important away from the low-energy region \cite{KarstenSmit1981NPB}. These terms break the parity symmetry that is locally preserved at the Fermi surface and thereby fix the anomalous Hall response \cite{Fu2024NatCommun,WanSun2025PRB}. In this sense, the half-quantized Hall conductance of the ordinary Dirac parity anomaly originates from the coexistence of a parity-invariant massless Dirac cone at low energy and a parity-breaking lattice regularization at high energy. It does not come from assigning a half Chern number to an isolated massive Dirac cone \cite{Zou2023PRB,Shen2026PRB}.

The right part of Fig.~\ref{fig:fig1} raises the central question of this work: does the same lattice-regularized logic produce a higher-order generalization of the parity-anomalous Hall response when the gapless point is replaced by a chiral Lifshitz fermion with low-energy coupling $(q_x+\ii q_y)^N$? Compared with an ordinary linear Dirac cone, this fermion has a nonlinear dispersion, $E\sim q^N$. The more important change is geometric. For the linear cone, the in-plane pseudospin winds once as momentum travels around a Fermi contour enclosing the gapless point. This unit winding produces the Fermi-surface Berry phase $\pi$ behind the half-quantized response $\sigma_H=e^2/2h$ marked in the left part of Fig.~\ref{fig:fig1} \cite{Berry1984PRS,Haldane2004PRL,XiaoChangNiu2010RMP}. For a chiral Lifshitz fermion, the pseudospin instead winds $N$ times, while the local parity symmetry on the Fermi surface is still preserved. Whether this $N$-fold winding survives lattice regularization, and what Hall response it enforces, {\color{black}is the central question illustrated} in the right part of Fig.~\ref{fig:fig1}. In the rest of this section, Sec.~\ref{subsec:fermi-loop} formalizes the parity-invariant Fermi loop of a chiral Lifshitz fermion, Sec.~\ref{subsec:lattice-cone} shows that any lattice completion necessarily breaks parity at high energies, and Sec.~\ref{subsec:n-half-derivation} derives the resulting winding-controlled Hall coefficient.

\subsection{Chiral Lifshitz Fermion and Fermi-Loop Winding}
\label{subsec:fermi-loop}

{\color{black}We consider a massless chiral Lifshitz fermion of order $N$.} Around the band-touching point, we write its two-band Hamiltonian as
\begin{equation}
h_N^{\mathrm{low}}(\mathbf{q})=
\epsilon_0(\mathbf{q})\sigma_0
+\lambda_N q^N
\left[
\cos(N\theta)\sigma_x+\sin(N\theta)\sigma_y
\right],
\label{eq:HN-low}
\end{equation}
where $\mathbf{q}=q(\cos\theta,\sin\theta)$ is measured from the touching point, $\sigma_0$ is the identity, $\sigma_{x,y}$ are Pauli matrices acting on the two-band pseudospin space, and $\lambda_N$ is a real constant setting the energy scale of the $N$th-order band touching. {\color{black}We denote the coefficients of $\sigma_x$ and $\sigma_y$ by $d_x(\mathbf q)=\lambda_Nq^N\cos(N\theta)$ and $d_y(\mathbf q)=\lambda_Nq^N\sin(N\theta)$, respectively.}
{\color{black}
\hypersetup{linkcolor=blue}
The $N\theta$ dependence in Eq.~\eqref{eq:HN-low} is not an independent assumption, but follows directly from the defining chiral coupling of an $N$th-order Lifshitz fermion,

\begin{equation*}
d_x+i d_y\propto(q_x+i q_y)^N=q^Ne^{iN\theta}.
\end{equation*}

Thus, the same low-energy structure gives both the nonlinear dispersion $E\propto q^N$ and the $N$-fold pseudospin winding. An $N$th-order dispersion by itself does not imply this winding: replacing $N\theta$ by $\theta$ would leave the spectrum unchanged but give $W=1$, and hence a winding-controlled anomalous contribution $e^2/2h$.} No mass term $m\sigma_z$ is included, since the cone is massless by construction. The scalar term $\epsilon_0(\mathbf{q})$ is taken to be parity even, $\epsilon_0(q_x,-q_y)=\epsilon_0(q_x,q_y)$, and allows for particle-hole asymmetry. Its value at the touching point only fixes the energy origin and can be subtracted when convenient. Since this term shifts both bands equally, it affects neither the eigenstates nor the Berry phases below, but only the shape of the constant-energy contours.

The spectrum of the massless chiral Lifshitz fermion is
\begin{equation}
E_{\pm}(\mathbf{q})=\epsilon_0(\mathbf{q})\pm |\lambda_N|q^N ,
\label{eq:HN-spectrum}
\end{equation}
an $N$th-order band touching with dynamical scaling $E\sim q^N$. When the Fermi energy lies inside the low-energy window, slightly away from the touching point, the Fermi surface is a small loop $\mathcal{C}$ encircling the node. On this loop the in-plane pseudospin angle equals $N\theta$, so the pseudospin winds $N$ times as $\mathbf{q}$ travels once around $\mathcal{C}$. This winding is the chirality that distinguishes the chiral Lifshitz fermion from a merely nonlinear band crossing.

The Fermi loop of this massless node is parity invariant in the following sense. In two spatial dimensions, parity acts as the mirror transformation $(q_x,q_y)\rightarrow(q_x,-q_y)$, accompanied by the internal operation $\mathcal{P}=\sigma_x$. Under $q_y\rightarrow-q_y$ one has $\theta\rightarrow-\theta$, so the $\sigma_x$ component of Eq.~\eqref{eq:HN-low} is even while the $\sigma_y$ component is odd, and therefore
\begin{equation}
\mathcal{P}h_N^{\mathrm{low}}(q_x,q_y)\mathcal{P}^{-1}
=h_N^{\mathrm{low}}(q_x,-q_y).
\label{eq:parity-low-energy}
\end{equation}
Equation~\eqref{eq:parity-low-energy} shows that an equal-energy Fermi loop encircling the node is mapped back to itself under parity. The massless chiral Lifshitz fermion is therefore the higher-order counterpart of the parity-invariant massless Dirac cone: the single pseudospin winding of the linear Dirac cone is replaced by an $N$-fold pseudospin winding, while the local Fermi loop remains parity symmetric. The remaining questions are how such a parity-invariant chiral Lifshitz fermion can be embedded in a lattice model, how fermion doubling and regularization are handled, and what parity-anomalous Hall response follows.

\subsection{Fermion Doubling and Parity-Breaking Lattice Regularization}
\label{subsec:lattice-cone}

We now ask how the parity-invariant chiral Lifshitz fermion described above
can be embedded in a lattice model. The relevant topological charge of the
isolated node is the vorticity of
\[
f(\mathbf q)\equiv d_x(\mathbf q)+i d_y(\mathbf q),
\]
defined on a small loop \(\mathcal C\) enclosing the touching point:
\[
W=\frac{1}{2\pi}\oint_{\mathcal C}d\arg f(\mathbf q)=N .
\]
This winding is a local property of the low-energy node. On a lattice,
however, \(f(\mathbf k)\) is a periodic function on the Brillouin torus. For
a globally defined two-band texture with isolated zeros, the total vorticity
must vanish,
\[
\sum_a W_a=0,\qquad
W_a=\frac{1}{2\pi}\oint_{\mathcal C_a}d\arg f(\mathbf k).
\]
Therefore a target low-energy node with \(W=N\) must be accompanied by
compensating high-energy zeros whose total winding is \(-N\). These
compensating zeros are the higher-order analogue of fermion doublers.

{\color{black}To retain a single low-energy \(N\)-fold Lifshitz node},
rather than resolving it into \(N\) separated Dirac cones, we use a
winding-preserving lattice completion. A convenient representative is
\[
H_{\rm lat}^{(N)}(\mathbf k)
=
\operatorname{Re} f_N(\mathbf k)\,\sigma_x
+
\operatorname{Im} f_N(\mathbf k)\,\sigma_y
+
M_{\rm reg}(\mathbf k)\sigma_z ,
\]
with $f_N(\mathbf k)=\lambda_N(\sin k_x+i\sin k_y)^N .
$
The zeros of \(f_N\) occur at
$
(0,0),\quad (\pi,0),\quad (0,\pi),\quad (\pi,\pi),
$ with vorticities
$+N,\quad -N,\quad -N,\quad +N ,
$
respectively, so the total vorticity vanishes as required.

The regulator mass \(M_{\rm reg}\) is chosen to gap the compensating zeros
without modifying the leading low-energy Lifshitz Hamiltonian at the target
valley. Near \(\mathbf k=0\), we require
\[
M_{\rm reg}(\mathbf q)=o(|\mathbf q|^N),
\]
while \(M_{\rm reg}\) is finite at the high-energy zeros. One explicit choice is
\[
M_{\rm reg}(\mathbf k)
=
m_{\rm W}
\left(2-\cos k_x-\cos k_y\right)^r,
\qquad 2r>N .
\]
This mass vanishes to order \(2r\) near \((0,0)\), but gaps the zeros at
\((\pi,0)\), \((0,\pi)\), and \((\pi,\pi)\).

The role of \(M_{\rm reg}\) is analogous to a Wilson mass. Since the local
parity operation of the isolated fermion is represented by
\(\mathcal P=\sigma_x\), a constant \(\sigma_z\) mass is parity odd,
\[
\mathcal P\sigma_z\mathcal P^{-1}=-\sigma_z .
\]
Thus the regulator leaves the target massless fermion parity invariant near the
target node, but gaps the compensating doublers through parity-breaking mass
terms at high-energy momenta. In this sense, a single low-energy chiral Lifshitz
fermion can be isolated only after a parity-breaking lattice regularization is
specified.

\subsection{Anomalous Hall Response of Chiral Lifshitz Fermions}
\label{subsec:n-half-derivation}

Having separated the parity-invariant chiral Lifshitz fermion from the high-energy regularization, we now extract the Hall response tied to the low-energy sector.
The relevant quantity is not the nonlinear dispersion by itself, but the phase winding of the in-plane pseudospin around a small constant-energy contour $\mathcal{C}$ enclosing the massless node. In a lattice completion this winding is defined as
\begin{equation}
W=
\frac{1}{2\pi}
\oint_{\mathcal{C}}
\dd\arg\left[d_x(\mathbf{k})+\ii d_y(\mathbf{k})\right].
\label{eq:wfs}
\end{equation}
Near the target node, $d_x+\ii d_y\simeq\lambda_N(q_x+\ii q_y)^N$, and therefore $W=N$. {\color{black}The same definition applies to the in-plane Pauli-matrix coefficients of a microscopic Hamiltonian projected onto its low-energy two-band subspace.}

This winding directly fixes the Berry phase. For $\lambda_N>0$, a convenient eigenstate of the upper band of Eq.~\eqref{eq:HN-low} is
\begin{equation}
\ket{u_+(\theta)}=
\frac{1}{\sqrt{2}}
\begin{pmatrix}
1\\
\e^{\ii N\theta}
\end{pmatrix}.
\label{eq:eigenstate}
\end{equation}
Using the Berry connection $\mathbf{A}=\ii\langle u_+|\nabla_{\mathbf{k}}u_+\rangle$ \cite{Berry1984PRS,XiaoChangNiu2010RMP}, the Berry phase accumulated along $\mathcal{C}$ is
\begin{equation}
\gamma=
\oint_{\mathcal{C}}
\mathbf{A}\cdot\dd\mathbf{k}
=\pm \pi W
=\pm N\pi ,
\label{eq:berry-phase}
\end{equation}
where the sign depends on the orientation of $\mathcal{C}$, the band convention, and the chirality of the fermion.

When the Fermi energy lies inside the low-energy window and slightly away from the touching point, it cuts one of the gapless bands of the chiral Lifshitz fermion and forms a small Fermi loop $\mathcal C$ around the node. The contribution of this partially filled band to the intrinsic Hall response can be written as a Berry-curvature integral over its occupied region. Away from the nodal singularity, the Berry curvature is locally the curl of the Berry connection $\mathbf A$, so Stokes' theorem reduces the low-energy contribution to a Berry-connection line integral on the Fermi loop \cite{Haldane2004PRL,XiaoChangNiu2010RMP}. The anomalous Hall response associated with the massless chiral Lifshitz fermion is therefore
\begin{equation}
\sigma_{xy}^{\mathrm{anom}}
=
\frac{e^2}{h}
\oint_{\mathcal{C}}
\frac{\dd\mathbf{k}}{2\pi}\cdot\mathbf{A}
=
\frac{W}{2}\frac{e^2}{h}.
\label{eq:n-half-anom}
\end{equation}
The important point is that $\mathcal C$ is the Fermi loop of a partially filled gapless band; Eq.~\eqref{eq:n-half-anom} therefore gives the low-energy Hall contribution from this partially filled band. This result follows from the Berry phase on the Fermi loop: as long as the loop encloses the same massless chiral Lifshitz fermion and carries the same winding, the line integral gives the same $W/2$ coefficient \cite{Zou2023PRB,Shen2026PRB}. In a full lattice model, completely filled bands away from the node may still add an integer Hall conductance; the absolute Hall conductance is therefore fixed by the microscopic realization, whereas the $W/2$ term is the anomalous Hall response of the massless chiral Lifshitz fermion.

\section{Realization in Multilayer Haldane Models}
\label{sec:haldane-realization}

\subsection{Single Gapless Higher-Order Cone in Multilayer Haldane Models}
\label{subsec:single-gapless-cone}

The previous section derived the anomalous Hall response of a massless chiral Lifshitz fermion from the Fermi-loop winding. We now test this result in a concrete lattice realization. A natural starting point is the Haldane model, where tuning the sublattice potential to a critical value closes the gap at one valley while keeping the other valley massive, thereby realizing a single massless Dirac cone with the required lattice regularization \cite{Haldane1988,Zou2022PRB,Zou2023PRB}. To extend this setting beyond the conventional linear cone, we consider chiral multilayer Haldane models \cite{WuZhangZubkov2022}. As in chiral multilayer graphene, interlayer dimerization can convert the low-energy single-valley Dirac dispersion into a chiral Lifshitz fermion, whose effective order and pseudospin winding number are determined by the stacking sequence \cite{MinMacDonald2008,KoshinoMcCann2009,McCannKoshino2013,ZhaoZhangZhang2025}. Therefore, multilayer Haldane models provide an ideal lattice platform for testing the anomalous Hall response of chiral Lifshitz fermions: they retain the valley-selective regularization mechanism of the original Haldane model, while the stacking sequence and layer number provide control parameters for tuning the gapless fermion and its Fermi-loop winding number.

\begin{figure}[htbp]
    \includegraphics[width=\linewidth]{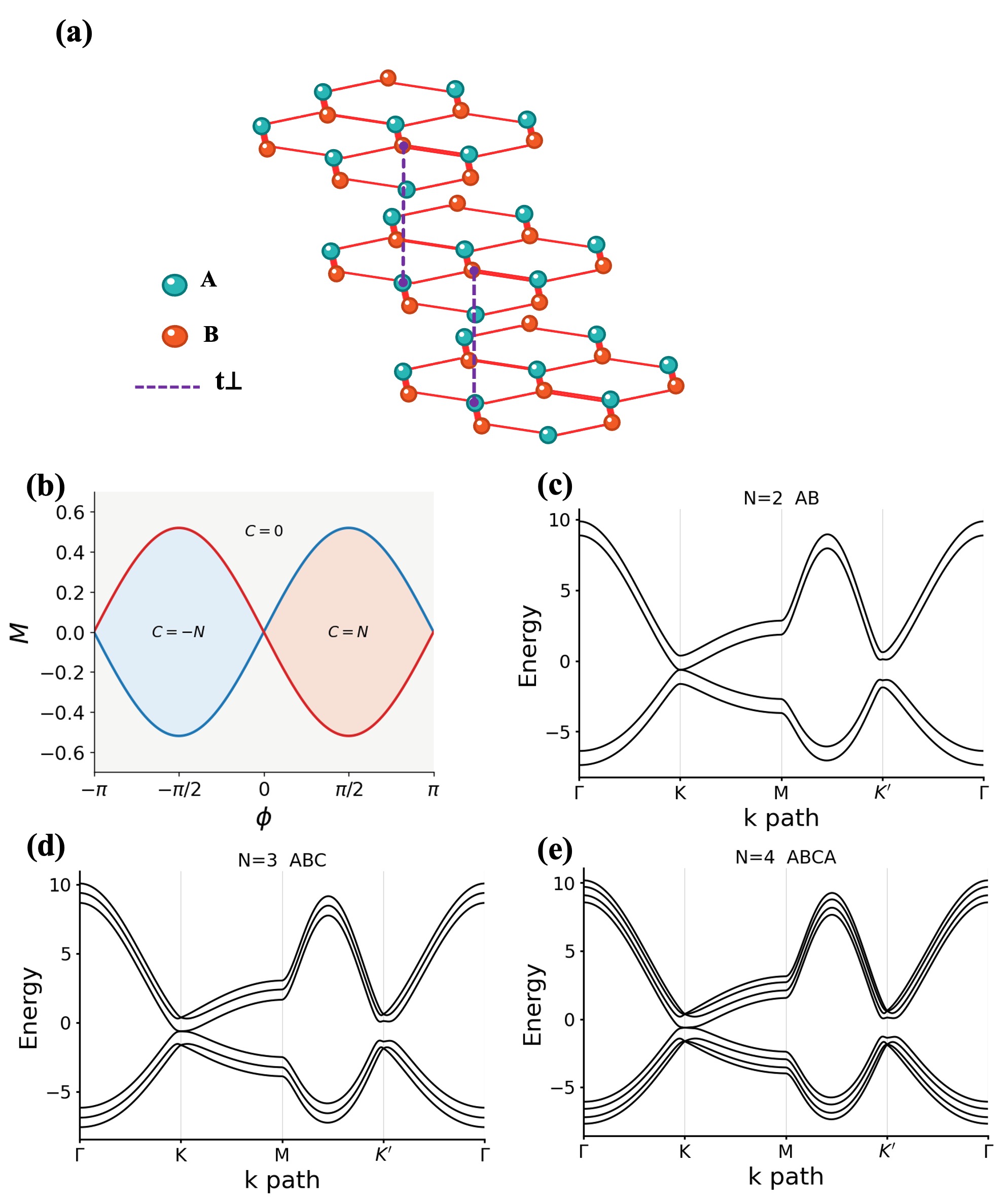}
    \caption{Multilayer Haldane model. (a) Stacked honeycomb lattice; the two colors denote the $A$ and $B$ sublattices, and the purple dashed bonds denote the interlayer dimer hopping $\tp$. (b) Phase diagram of the $N$-layer model in the $(\phi,M)$ plane; the critical lines $M=\pm\sqrt{3}\,t_2\sin\phi$ separate the $C=0$ and $C=\pm N$ phases. (c)--(e) Bulk spectra along $\Gamma$-$K$-$M$-$K'$-$\Gamma$ for the AB $(N=2)$, ABC $(N=3)$, and ABCA $(N=4)$ stacks at $\phi=0.8$ and $M=\sqrt{3}\,t_2\sin\phi$, where the middle gap closes at the $K$ valley while the $K'$ valley remains gapped.}
    \label{fig:model-band}
\end{figure}

Figure~\ref{fig:model-band}(a) shows the multilayer Haldane lattice considered below. Each layer is a honeycomb lattice with two sublattices, where the green and red sites denote the $A$ and $B$ sublattices, respectively. The solid bonds represent the intralayer nearest-neighbor hopping on each honeycomb layer. The purple dashed bonds denote the interlayer dimer hopping, which couples selected $B_\ell$ sites in layer $\ell$ to $A_{\ell+1}$ sites in the adjacent layer. Different choices of the in-plane displacement of these dimer bonds generate the AB, ABC, and ABCA stacking sequences considered here.

The corresponding $N$-layer Hamiltonian is
\begin{equation}
H_N=\sum_{\ell=1}^{N}H_{\mathrm{H}}^{(\ell)}+H_{\perp},
\label{eq:multi-haldane}
\end{equation}
with
\begin{align}
H_{\mathrm{H}}^{(\ell)}
={}&
t_1\sum_{\langle ij\rangle}
\left(c_{\ell i}^{\dagger}c_{\ell j}+\mathrm{H.c.}\right)
\nonumber\\
&+
t_2\sum_{\langle\!\langle ij\rangle\!\rangle}
\left(
\e^{\ii\nu_{ij}\phi}
c_{\ell i}^{\dagger}c_{\ell j}
+\mathrm{H.c.}
\right)
\nonumber\\
&+
M\sum_i \xi_i c_{\ell i}^{\dagger}c_{\ell i},
\label{eq:single-layer-haldane-real-space}
\\
H_{\perp}
={}&
\tp\sum_{\ell=1}^{N-1}\sum_{\mathbf{R}}
\left(
c_{\ell,B,\mathbf{R}}^{\dagger}
c_{\ell+1,A,\mathbf{R}+\boldsymbol{\delta}_{\ell}}
+\mathrm{H.c.}
\right).
\label{eq:interlayer-hopping}
\end{align}
Here $c_{\ell i}^{\dagger}$ creates an electron on site $i$ in layer $\ell$. The parameter $t_1$ denotes the intralayer nearest-neighbor hopping amplitude, while $t_2$ is the amplitude of the complex next-nearest-neighbor Haldane hopping. The phase $\phi$, together with the orientation factor $\nu_{ij}=\pm1$, encodes the staggered flux pattern of the Haldane model. The last term is the Semenoff sublattice potential, with $\xi_i=+1$ on the $A$ sublattice and $\xi_i=-1$ on the $B$ sublattice. The interlayer term $H_{\perp}$ describes the dimer hopping $\tp$ shown by the purple dashed bonds in Fig.~\ref{fig:model-band}(a), where the layer-dependent displacement $\boldsymbol{\delta}_{\ell}$ specifies the stacking pattern. In this way, the same Hamiltonian form describes the AB, ABC, and ABCA multilayer structures \cite{WuZhangZubkov2022}. {\color{black}Throughout the numerical calculations and figures, energies are measured in units of the interlayer hopping, with $\tp=1$. We use $t_1=2.7$ and $t_2=0.3$.}

To obtain the band structure, we Fourier transform the real-space tight-binding Hamiltonian in Eqs.~\eqref{eq:multi-haldane}--\eqref{eq:interlayer-hopping} into momentum space. This gives a $2N\times2N$ Bloch Hamiltonian with a block-tridiagonal structure in the layer index: the diagonal blocks are the monolayer Haldane Hamiltonians, while the off-diagonal blocks couple only the dimer pair $B_\ell$ and $A_{\ell+1}$. This form also provides the starting point for identifying the low-energy subspace used in the winding-number calculation below \cite{ZhaoZhangZhang2025}. In the Bloch basis $\Psi_{\mathbf{k}}$ ordered as $(1A,1B,2A,2B,\ldots,NA,NB)$, the Hamiltonian is
\begin{equation}
\resizebox{\columnwidth}{!}{$\displaystyle
H_N(\mathbf{k})=
\begin{pmatrix}
g_{\mathbf{k}}^+ & f_{\mathbf{k}} & 0 & 0 & 0 & \cdots & 0 & 0\\
f_{\mathbf{k}}^* & g_{\mathbf{k}}^- & T_1(\mathbf{k}) & 0 & 0 & \cdots & 0 & 0\\
0 & T_1^*(\mathbf{k}) & g_{\mathbf{k}}^+ & f_{\mathbf{k}} & 0 & \cdots & 0 & 0\\
0 & 0 & f_{\mathbf{k}}^* & g_{\mathbf{k}}^- & T_2(\mathbf{k}) & \cdots & 0 & 0\\
0 & 0 & 0 & T_2^*(\mathbf{k}) & g_{\mathbf{k}}^+ & \cdots & 0 & 0\\
\vdots & \vdots & \vdots & \vdots & \vdots & \ddots & \vdots & \vdots\\
0 & 0 & 0 & 0 & 0 & \cdots & g_{\mathbf{k}}^+ & f_{\mathbf{k}}\\
0 & 0 & 0 & 0 & 0 & \cdots & f_{\mathbf{k}}^* & g_{\mathbf{k}}^-
\end{pmatrix}.
$}
\label{eq:bloch-matrix}
\end{equation}
The matrix elements are
\begin{equation}
\begin{aligned}
g_{\mathbf{k}}^\pm
&=
\pm M
+2t_2\sum_{\boldsymbol{b}}
\cos\left(\mathbf{k}\cdot\boldsymbol{b}\pm\phi\right),\\
f_{\mathbf{k}}
&=
t_1\sum_{\boldsymbol{a}}
\e^{\ii\mathbf{k}\cdot\boldsymbol{a}},\\
T_{\ell}(\mathbf{k})
&=
\tp\,\e^{\ii\mathbf{k}\cdot\boldsymbol{\delta}_{\ell}} .
\end{aligned}
\label{eq:bloch-elements}
\end{equation}
Here the lattice constant is set to unity, with primitive vectors $\boldsymbol{a}_1=(1,0)$ and $\boldsymbol{a}_2=(1/2,\sqrt{3}/2)$. The sum over $\boldsymbol{a}$ runs over the three nearest-neighbor vectors connecting an $A$ site to its $B$ neighbors, while the sum over $\boldsymbol{b}$ runs over the oriented next-nearest-neighbor vectors $\boldsymbol{b}\in\{-\boldsymbol{a}_1,-\boldsymbol{a}_2,\boldsymbol{a}_1-\boldsymbol{a}_2\}$, along which $\nu_{ij}=+1$ on the $A$ sublattice and $-1$ on the $B$ sublattice. Here $T_\ell$ couples the dimer pair $B_\ell$ and $A_{\ell+1}$. The in-plane Bravais offset $\boldsymbol{\delta}_\ell$ determines the phase of $T_\ell(\mathbf{k})$, and the sequence $\{\boldsymbol{\delta}_\ell\}$ specifies the AB, ABC, or ABCA stacking. Thus $g_{\mathbf{k}}^\pm$ contains the sublattice mass and Haldane flux on each layer, $f_{\mathbf{k}}$ gives the intralayer honeycomb hopping, and $T_\ell(\mathbf{k})$ carries the stacking-dependent interlayer phase.

Figure~\ref{fig:model-band}(b) summarizes the phase diagram of the $N$-layer Haldane model in the $(\phi,M)$ plane. The phase boundaries are fixed by the valley-resolved masses inherited from the Haldane layers, $m_{K,K'}=M\mp \sqrt{3}\,t_2\sin\phi$, and therefore occur at $M=\pm \sqrt{3}\,t_2\sin\phi$. Along either boundary, the gap closes at one valley while the other valley remains massive. Away from these critical lines, the system is fully gapped, and its total Chern number can be computed over the occupied bands using the Fukui--Hatsugai--Suzuki lattice-gauge method \cite{Thouless1982PRL,Fukui2005FHS},
\begin{equation}
C=
\sum_{n\in\mathrm{occ}}
\frac{1}{2\pi\ii}
\sum_{\mathbf{k}}
\ln
\frac{
U_x^n(\mathbf{k})U_y^n(\mathbf{k}+\delta k_x)
}{
U_y^n(\mathbf{k})U_x^n(\mathbf{k}+\delta k_y)
},
\end{equation}
where
\begin{equation}
U_{x/y}^n(\mathbf{k})
=
\frac{
\langle u^n(\mathbf{k})|u^n(\mathbf{k}+\delta k_{x/y})\rangle
}{
|\langle u^n(\mathbf{k})|u^n(\mathbf{k}+\delta k_{x/y})\rangle|
}.
\end{equation}
Here $n$ runs over the occupied bands. The resulting gapped phases have $C=0$ in the trivial regions and $C=\pm N$ in the topologically nontrivial regions, showing that the Chern number is multiplied by the number of layers. The phase boundary itself is gapless, so $C$ is not defined there; rather, the Chern number changes by $\Delta C=\pm N$ across the boundary. In the following, we tune $M$ to one of these critical lines, so that a single valley becomes gapless while the other valley remains gapped. This produces a valley-selective low-energy gapless sector, while the gapped opposite valley and high-energy bands provide the lattice regularization. In this work, we take $\phi=0.8$ and tune to the $K$-valley boundary, $M=\sqrt{3}t_2\sin\phi\simeq0.37275$.

Figures~\ref{fig:model-band}(c)--\ref{fig:model-band}(e) show the corresponding bulk band structures for the AB, ABC, and ABCA stacks, with $N=2$, $3$, and $4$, respectively. In all three cases, the middle gap closes at the $K$ valley, while the $K'$ valley remains gapped. The gapless point lies at $E_K\simeq -0.62704$, and the remaining $K'$ valley gap is finite, $\Delta_{K'}\simeq1.491$. Thus the spectra explicitly realize the valley-selective critical point indicated by Fig.~\ref{fig:model-band}(b). At the same time, the shape of the gapless cone at $K$ evolves visibly as the number of layers increases. Tuning to the phase boundary therefore produces an isolated single-valley gapless critical point on the lattice. We next identify this critical point as a chiral Lifshitz fermion by analyzing its dispersion and Fermi-loop winding.

\subsection{Chiral Lifshitz Fermions in Multilayer Haldane Models}
\label{subsec:winding-number}

The band structures in Fig.~\ref{fig:model-band} show that the critical point is valley selective: the middle gap closes at the $K$ valley while the opposite valley remains gapped. This provides the lattice-regularized single-valley setting needed for the anomalous Hall response, but the gap closing alone does not determine its coefficient. As shown in Sec.~\ref{sec:higher-order-theory}, the coefficient is fixed by the Fermi-loop winding of the parity-invariant low-energy fermion. We therefore analyze the dispersion and projected pseudospin texture near the closing point to determine whether it realizes an $N$th-order chiral Lifshitz fermion.

\begin{figure}[htbp]
    \centering
    \includegraphics[width=\linewidth]{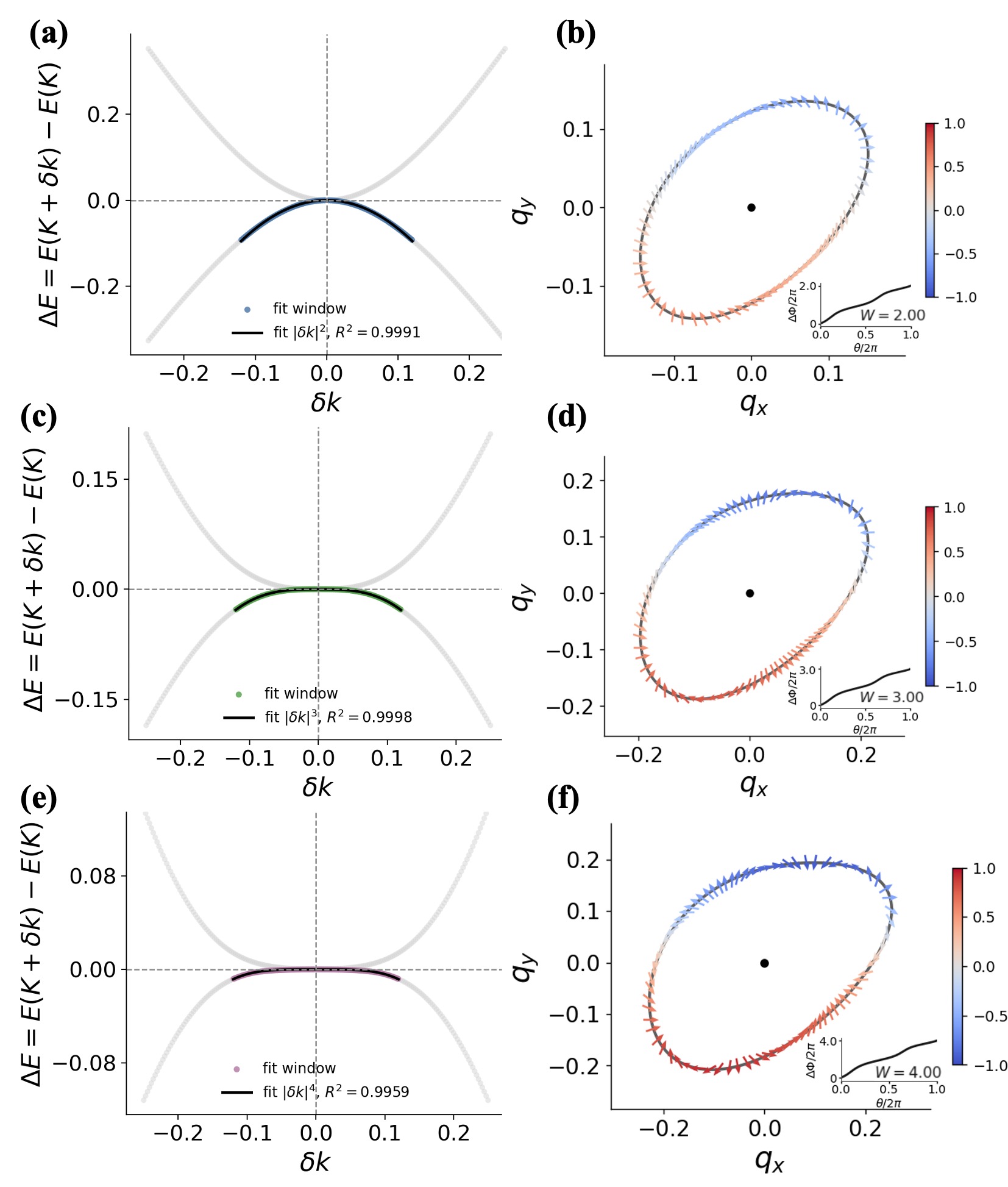}
    \caption{Dispersion and winding number of the gapless cone at the $K$ valley for the AB (top row), ABC (middle row), and ABCA (bottom row) stacks. (a), (c), (e) Power-law fits $\Delta E\propto|\delta k|^{n}$ of the middle-band dispersion within the fit window, giving $n=2$, $3$, and $4$. (b), (d), (f) Projected pseudospin textures $\mathbf{d}'(\mathbf{q})$ on the constant-energy contours $\mathcal{C}_K$: the arrows show the in-plane components $(d'_x,d'_y)$ and the color shows $d'_z$. The insets give the unwrapped phase of $d'_x+\ii d'_y$ along the contour, yielding $W=2$, $3$, and $4$ from Eq.~\eqref{eq:numerical-winding}.}
    \label{fig:winding}
\end{figure}

We first examine the dispersion in the immediate neighborhood of this critical valley. At the tuned mass, the gap at $K=(4\pi/3,2\pi/3)$ is numerically closed, $\Delta_K\simeq4.4\times10^{-9}$. We take a local radial cut $\mathbf{k}=K+\delta k\,\hat{\mathbf{e}}_1$, where $\hat{\mathbf{e}}_1$ denotes the positive $k_1$ direction, equivalently the reciprocal direction dual to $\boldsymbol{a}_1$, and fit the middle valence-band shift
\begin{equation}
\Delta E(\delta k)=E(K+\delta k)-E(K)\propto|\delta k|^n .
\label{eq:power-fit}
\end{equation}
Within the fitting window $|\delta k|\leq0.12$, the AB stack is well described by a quadratic dispersion in Fig.~\ref{fig:winding}(a), the ABC stack by a cubic dispersion in Fig.~\ref{fig:winding}(c), and the ABCA stack by a quartic dispersion in Fig.~\ref{fig:winding}(e). The corresponding $R^2$ values are $0.9991$, $0.9998$, and $0.9959$, respectively. Thus, for the three stacking sequences studied here, the single-valley cone evolves from quadratic to cubic and quartic near $K$.

We next extract the winding number from the low-energy subspace near the same Fermi point. In the Bloch matrix of Eq.~\eqref{eq:bloch-matrix}, the stacking-dependent elements $T_\ell(\mathbf{k})$ hybridize $B_\ell$ and $A_{\ell+1}$ into dimerized states. The two outer orbitals $A_1$ and $B_N$ are not involved in these interlayer dimers and therefore carry the low-energy spectral weight near the gapless valley. We choose the low-energy sector as $L=(A_1,B_N)$, with basis $\Psi_{\mathrm{low}}=(c_{A_1},c_{B_N})^T$, and collect the remaining orbitals into $D=(B_1,A_2,B_2,A_3,\ldots,B_{N-1},A_N)$. Reordering the Bloch Hamiltonian in the basis $(L,D)$ gives
\begin{equation}
\begin{aligned}
H_N(K+\mathbf{q})
=
\begin{pmatrix}
H_L(\mathbf{q}) & V(\mathbf{q})\\
V^\dagger(\mathbf{q}) & H_D(\mathbf{q})
\end{pmatrix}.
\end{aligned}
\label{eq:block-decomposition}
\end{equation}
The subspace $D$ is then eliminated by the Schur-complement downfolding, giving
\begin{equation}
\begin{aligned}
H_{\mathrm{low}}^{(K)}(\mathbf{q})
={}&
H_L(\mathbf{q})
-
V(\mathbf{q})
\left[
H_D(\mathbf{q})-E_K I_D
\right]^{-1}
V^\dagger(\mathbf{q})
\\
={}&
d'_0(\mathbf{q})\sigma_0
+d'_x(\mathbf{q})\sigma_x
+d'_y(\mathbf{q})\sigma_y
+d'_z(\mathbf{q})\sigma_z ,
\end{aligned}
\label{eq:low-projection}
\end{equation}
where $\mathbf{q}=\mathbf{k}-K$. The blocks $H_L$, $H_D$, and $V$ denote the Hamiltonian within the low-energy subspace, within the eliminated subspace, and between the two subspaces, respectively, and $I_D$ is the identity matrix in the eliminated subspace $D$. The second term in Eq.~\eqref{eq:low-projection} retains the virtual hopping processes through $D$. Here $\sigma_0$ is the identity matrix and $\sigma_{x,y,z}$ are Pauli matrices acting on the low-energy orbital space $(A_1,B_N)$. The coefficients $d'_x$ and $d'_y$ define the in-plane pseudospin components. Therefore, the phase of $d'_x+\ii d'_y$ gives the in-plane pseudospin angle used to extract the winding number.
At the gapless $K$ valley, we characterize the cone by taking a small closed momentum contour $\mathcal{C}_K$ around the band-touching point. In this work, $\mathcal{C}_K$ is chosen as the constant-energy loop at an energy offset $|E-E_K|=0.1$ from the gapless energy $E_K$, within the low-energy region where the projected two-band description remains valid. The winding number is defined as the winding of the in-plane pseudospin phase,
\begin{equation}
W
=
\frac{1}{2\pi}
\oint_{\mathcal{C}_K}
\dd \arg\left[
d'_x(\mathbf{q})+\ii d'_y(\mathbf{q})
\right].
\label{eq:numerical-winding}
\end{equation}
Numerically, we evaluate $d'_x+\ii d'_y$ along $\mathcal{C}_K$, unwrap its phase, and divide the total phase advance by $2\pi$.

The $\mathbf d'(\mathbf q)$ textures are shown in Figs.~\ref{fig:winding}(b), (d), and (f). The vector field represents the in-plane components $d'_x$ and $d'_y$, while the color represents the $d'_z$ component. The insets show the unwrapped phase $\arg(d'_x+\ii d'_y)$ along the contour. For the AB stack, Fig.~\ref{fig:winding}(b) shows two full rotations and gives $W=2$. For the ABC stack, Fig.~\ref{fig:winding}(d) shows three rotations and gives $W=3$. For the ABCA stack, Fig.~\ref{fig:winding}(f) shows four rotations and gives $W=4$. Therefore, within these stacking sequences, increasing the layer number increases the winding number of the gapless cone.

For the AB, ABC, and ABCA stacks, the gap closing at the $K$ valley is therefore identified as an $N$th-order chiral Lifshitz fermion, with $N=2$, $3$, and $4$, respectively, and with the matching Fermi-loop winding $W=N$. The next section tests the Hall response associated with this low-energy structure in the full lattice model.

\subsection{Hall Response and Equilibrium-Current Signature}
\label{subsec:hall-current}

Having identified the winding number of the gapless fermion, we next test whether it produces the predicted Hall response in the full lattice model. This step goes beyond the projected two-band Hamiltonian near $K$ and includes the parity-breaking high-energy lattice regularization.
{\color{black}
\hypersetup{linkcolor=blue,citecolor=blue}
The current operator entering the Kubo response follows from the term
linear in the external vector potential in the Peierls-coupled lattice
Hamiltonian~\cite{Zou2023PRB,Fu2024NatCommun,Shen2026PRB}.} We compute $\sigma_{xy}(\EF)$ using the Kubo Berry-curvature formula \cite{cpp8-bgz5, s6pj-495v,XiaoChangNiu2010RMP,Nagaosa2010RMP},
\begin{equation}
\Omega_n(\mathbf{k})
=
-\sum_{m\neq n}
\frac{
2\,\mathrm{Im}\left[
\langle u_{n\mathbf{k}}|v_x|u_{m\mathbf{k}}\rangle
\langle u_{m\mathbf{k}}|v_y|u_{n\mathbf{k}}\rangle
\right]
}{
\left(E_{m\mathbf{k}}-E_{n\mathbf{k}}\right)^2
},
\label{eq:kubo-berry}
\end{equation}
where $|u_{n\mathbf{k}}\rangle$ is the Bloch eigenstate of the $n$th band with energy $E_{n\mathbf{k}}$, and {\color{black}$v_{x,y}=\partial H_N(\mathbf{k})/\partial k_{x,y}$ are the matrix-derivative vertices used here ($\hbar=1$ in the numerical implementation)}. The Hall conductance is obtained by integrating the occupied Berry curvature,
\begin{equation}
\sigma_{xy}(\EF)
=
-\frac{e^2}{h}
\sum_n
\int_{\mathrm{BZ}}
\frac{\dd^2 k}{2\pi}\,
f\!\left(E_{n\mathbf{k}}-\EF\right)
\Omega_n(\mathbf{k}) ,
\label{eq:kubo}
\end{equation}
where $\mathrm{BZ}$ denotes the first Brillouin zone and $f$ is the Fermi occupation function. This gives $\sigma_{xy}$ as a function of $\EF$, to be compared with the predicted value $\sigma_{xy}=W e^2/2h$.

\begin{figure}[htbp]
    \centering
    \includegraphics[width=\linewidth]{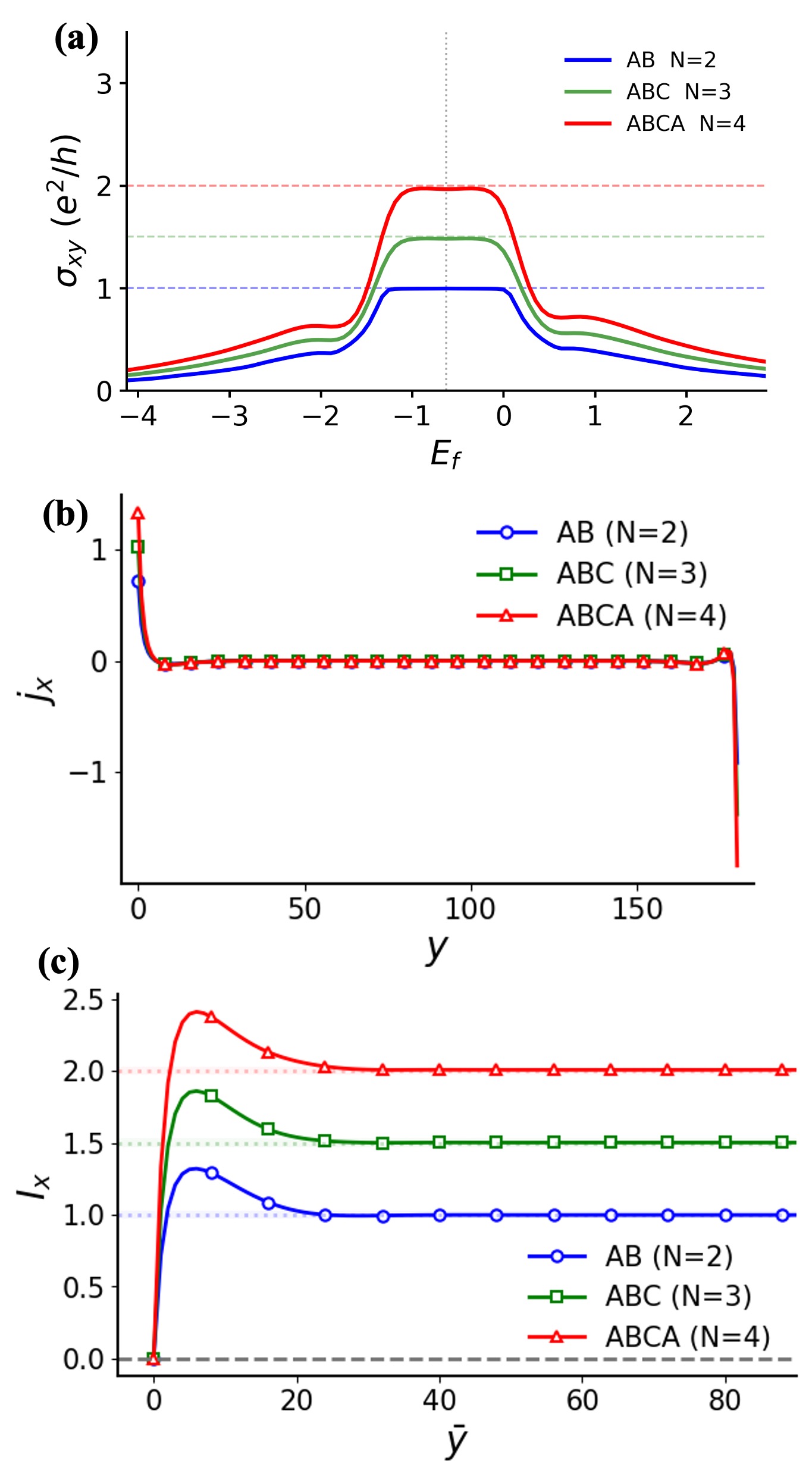}
    \caption{Hall response and equilibrium current. (a) Kubo Hall conductance $\sigma_{xy}(\EF)$ of the AB, ABC, and ABCA stacks, with plateaus at $1$, $3/2$, and $2$ in units of $e^2/h$; the vertical dotted line marks the gapless energy $E_K$. (b) Row-resolved equilibrium current density $j_x(y)$ at $E=-0.5$, calculated in a strip periodic along $x$ and of width $L_y=180$; the current is localized at the two boundaries. (c) Integrated current $I_x(E,\bar{y})$ at the same energy, which converges to $W/2=1$, $3/2$, and $2$ for the three stacks.}
    \label{fig:hall-current}
\end{figure}

As shown in Fig.~\ref{fig:hall-current}(a), the Kubo Hall conductance develops stable quantized plateaus inside the gap of the massive $K'$ valley. For the AB, ABC, and ABCA stacks, the plateau values are $\sigma_{xy}=1$, $3/2$, and $2$ in units of $e^2/h$, respectively. These values match the Fermi-loop prediction of the chiral Lifshitz fermions, $\sigma_{xy}=W e^2/2h$ for $W=2,3,4$.

The Hall plateau gives a momentum-space check of the winding-controlled response. We further calculate the fixed-energy equilibrium current in a strip geometry that is periodic along $x$ and finite along $y$. Following the Green-function construction used for parity-anomalous boundary currents \cite{Datta1995,Zou2022PRB,WanSun2025PRB,Liu2026PRB}, the row-resolved current is
\begin{equation}
j_x(E,y)
=
-\frac{e}{\pi h}
\int_{-\pi}^{\pi}\dd k_x\,
\mathrm{Im}\,
\mathrm{Tr}\left[
\langle y|v_x(k_x)G^R_{k_x}(E)|y\rangle
\right],
\label{eq:equilibrium-current}
\end{equation}
where $v_x(k_x)=\partial H(k_x)/\partial k_x$ and $G^R_{k_x}(E)=[E-H(k_x)+\ii\eta]^{-1}$, with {\color{black}numerical Lorentzian width $\eta=0.1$ in the above energy units}. Here $|y\rangle$ denotes the subspace of orbitals on the transverse row $y$ of the strip. Equivalently, $G^R_{k_x}(E;y,y')=\langle y|G^R_{k_x}(E)|y'\rangle$ denotes the row-resolved Green-function block connecting rows $y$ and $y'$. Once the row indices are fixed, the trace is taken only over the internal layer and sublattice degrees of freedom. {\color{black}The Smr\v{c}ka--St\v{r}eda decomposition of the bulk Hall conductivity is given in Appendix~\ref{app:streda}.}

Figure~\ref{fig:hall-current}(b) shows the distribution of the equilibrium current $j_x$ along the finite $y$ direction for different stacking sequences. The current is localized near the two boundaries and nearly vanishes in the bulk. As shown in Fig.~\ref{fig:hall-current}(c), the current flux is calculated from $I_x(E,\bar{y})=\int_0^{\bar{y}}\dd y\,j_x(E,y)$ with $\bar{y}<L_y/2$. As $\bar{y}$ increases, all stacks approach quantized plateaus: $I_x=1$ for AB, $I_x=3/2$ for ABC, and $I_x=2$ for ABCA. Similar to the half-quantized equilibrium current of a linear Dirac cone, these plateaus give the winding-controlled sequence $I_x=W/2$ in real space.

\subsection{Six-Terminal Transport under Dephasing}
\label{subsec:six-terminal}

\begin{figure}[htbp]
    \centering
    \includegraphics[width=\linewidth]{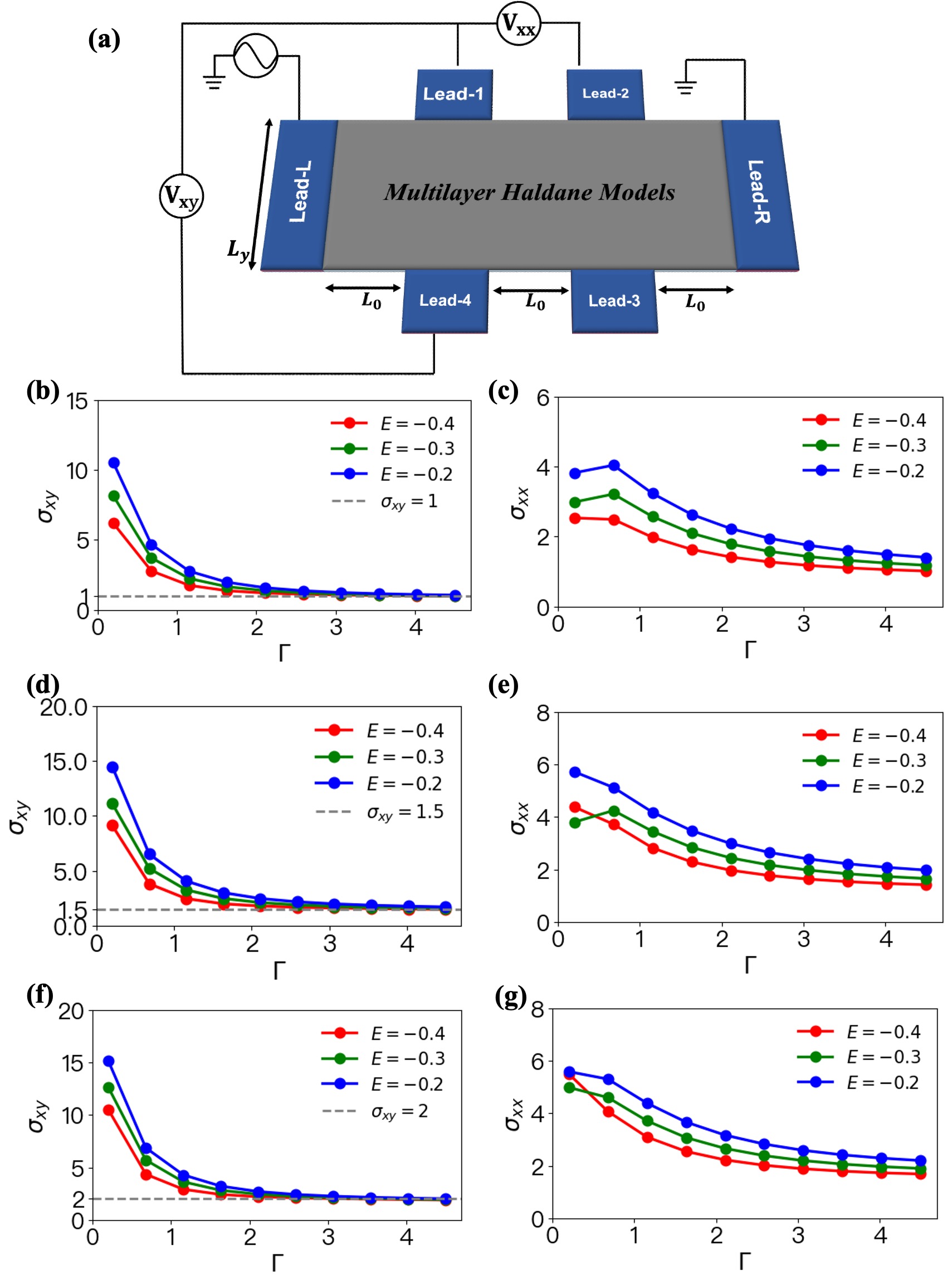}
    \caption{Dephasing-assisted six-terminal transport. (a) Hall-bar geometry: Lead-L and Lead-R are the source and drain, and Leads-1--4 are voltage probes attached at $x=L_x/3$ and $2L_x/3$, with $L_x=300$, $L_y=150$, and $L_0=L_x/3$. Each voltage probe contacts eight complete stack cells, the virtual-probe grid spacing is four unit cells, and the wide-band couplings are $\gamma_{\mathrm{lr}}=1.0$ and $\gamma_{\mathrm{probe}}=0.5$. (b), (d), (f) Hall conductivity $\sigma_{xy}$ versus dephasing strength $\Gamma$ for the AB, ABC, and ABCA stacks at $\EF=-0.4$, $-0.3$, and $-0.2$; the dashed lines mark $W/2=1$, $3/2$, and $2$. (c), (e), (g) Corresponding longitudinal conductivity $\sigma_{xx}$, which remains finite.}
    \label{fig:transport}
\end{figure}

To compute the transport response of the chiral Lifshitz fermions, we consider a six-terminal Hall-bar geometry and perform a standard transport simulation using nonequilibrium Green's functions \cite{Buttiker1986PRL,Buttiker1988IBM,Datta1995}.
As shown in Fig.~\ref{fig:transport}(a), Lead-L and Lead-R are the source and drain of the Hall bar, respectively, while Leads-1--4 act as voltage probes.
The scattering region is a rectangular flake cut from the multilayer Haldane lattice.
Orienting the zigzag direction of each honeycomb layer along $x$ and setting the lattice constant to unity, we retain every site of all $N$ layers whose position lies in the rectangle $0\leq x<L_x$, $0\leq y<(\sqrt{3}/2)L_y$, which spans $L_x\times L_y$ in-plane unit cells.
The two long edges at $y=0$ and $y\simeq(\sqrt{3}/2)L_y$ are zigzag terminated; since successive layers are displaced by the in-plane stacking shift, their outermost atomic rows are staggered from layer to layer, while all layers share the same rectangular envelope.
The source and drain contacts cover the full cross section at the two short ends, coupling to all sites, in every layer, within the outermost column at $x\simeq0$ or $x\simeq L_x$.
The four voltage probes are attached to the two zigzag edges at $x=L_x/3$ and $x=2L_x/3$.
Each probe contacts eight consecutive complete stack cells---a stack cell comprising the $2N$ sites $A_1,B_1,\ldots,A_N,B_N$ that share one in-plane unit cell---taken from the outermost row of complete cells along the corresponding edge, so that all layers are contacted {\color{black}equivalently}.
For Fig.~\ref{fig:transport}, we use $L_x=300$ and $L_y=150$, giving the probe separation $L_0=L_x/3=100$ and the effective width $W_{\mathrm{eff}}\simeq(\sqrt{3}/2)L_y$, defined as the transverse distance between the outermost retained atomic rows.
We introduce B\"uttiker virtual leads in the central region to simulate the dephasing effect \cite{Buttiker1986PRL,Buttiker1988IBM,Xing2008PRB,Zhou2022PRL}.
Electrons can enter these virtual leads and then return to the system with randomized phase, while the zero-current condition on every virtual lead prevents net particle loss.
Each virtual lead is attached to a single complete stack cell and couples to all of its $2N$ orbitals.
The dephasing cells form a regular grid covering the whole flake, with one cell selected every four unit cells along each primitive lattice direction; cells containing sites already coupled to one of the six real terminals are excluded.

The current of Lead-$p$ is determined by the Landauer--B\"uttiker formula \cite{Buttiker1986PRL,Buttiker1988IBM,Lambert1993,Datta1995,Xing2008PRB,5378-2fcs}
\begin{equation}
I_p=\frac{e^2}{h}
\sum_q T_{pq}\left(V_p-V_q\right),
\label{eq:lb}
\end{equation}
where $V_p$ is the voltage of Lead-$p$, and $p$ and $q$ run over both real and virtual leads. The transmission coefficient is given by $T_{pq}(E)=\mathrm{Tr}[\boldsymbol{\Gamma}_p(E)\mathbf{G}^R(E)\boldsymbol{\Gamma}_q(E)\mathbf{G}^A(E)]$, where $\mathbf{G}^R=(\mathbf{G}^A)^\dagger$ is the retarded Green's function. Using Dyson's equation, $\mathbf{G}^R(E)=[(E+\ii0^+)\mathbf{1}-\mathbf{H}_{\mathrm{cen}}-\sum_p\boldsymbol{\Sigma}_p^R(E)]^{-1}$. Here $\mathbf{H}_{\mathrm{cen}}$ is the isolated central stacked-Haldane Hamiltonian. $\boldsymbol{\Sigma}_p^R$ is the retarded self-energy from Lead-$p$, and $\boldsymbol{\Gamma}_p=\ii[\boldsymbol{\Sigma}_p^R-(\boldsymbol{\Sigma}_p^R)^\dagger]$ is the corresponding linewidth matrix. For the real leads, we use wide-band self-energies $\boldsymbol{\Sigma}_p^R=-\ii\gamma_p\mathbf{1}_p/2$, with $\gamma_p=\gamma_{\mathrm{lr}}$ for Lead-L and Lead-R and $\gamma_p=\gamma_{\mathrm{probe}}$ for Leads-1--4; the calculations in Fig.~\ref{fig:transport} use $\gamma_{\mathrm{lr}}=1.0$ and $\gamma_{\mathrm{probe}}=0.5$. For each virtual lead $\alpha$, we use $\boldsymbol{\Sigma}_{\alpha}^R=-\ii\Gamma\mathbf{1}_{\alpha}/2$, where $\Gamma$ is the dephasing strength. To simulate a Hall measurement, we set $V_{\mathrm L}=1$ and $V_{\mathrm R}=0$, and impose $I_1=I_2=I_3=I_4=I_{\alpha}=0$ for all virtual leads. Solving Eq.~\eqref{eq:lb} gives $I_{\mathrm L}=-I_{\mathrm R}$ and all terminal voltages.

To extract the longitudinal and Hall conductivities, we first obtain $R_{xy}=(V_1-V_4)/I_{\mathrm L}$ and $R_{xx}=(V_1-V_2)/I_{\mathrm L}$. With the geometrical factor $W_{\mathrm{eff}}/L_0$ defined by this finite Hall-bar shape, the resistivities and conductivities are calculated as
\begin{equation}
\begin{aligned}
\rho_{xy}&=R_{xy},&
\rho_{xx}&=\frac{W_{\mathrm{eff}}}{L_0}R_{xx},\\
\sigma_{xy}&=\frac{\rho_{xy}}{\rho_{xy}^2+\rho_{xx}^2},&
\sigma_{xx}&=\frac{\rho_{xx}}{\rho_{xy}^2+\rho_{xx}^2}.
\end{aligned}
\label{eq:rho-to-sigma}
\end{equation}

Figures~\ref{fig:transport}(b), \ref{fig:transport}(d), and \ref{fig:transport}(f) show the Hall conductivity $\sigma_{xy}$ of the AB, ABC, and ABCA stacks as a function of the dephasing strength $\Gamma$.
All the chosen Fermi energies cut the gapless cone at the $K$ valley and lie inside the gap of the massive $K'$ valley.
As $\Gamma$ increases, $\sigma_{xy}$ decays rapidly, and the curves for different Fermi energies converge: at strong dephasing they saturate at the plateaus $\sigma_{xy}\simeq1$, $3/2$, and $2$ (dashed lines) in units of $e^2/h$.
These stable plateau values agree with the winding numbers obtained above, as well as with the Kubo Hall conductivity and the equilibrium current.
The fact that the quantized plateau emerges only after sufficient dephasing is consistent with the interpretation that the quantization is carried by the local Hall response of the gapless chiral Lifshitz fermion rather than contributed by a massive Dirac cone \cite{Zhou2022PRL,WanSun2025PRB,Liu2026PRB}. It also shows that the Kubo and equilibrium-current plateaus survive in an open transport geometry.

Figures~\ref{fig:transport}(c), \ref{fig:transport}(e), and \ref{fig:transport}(g) show the corresponding longitudinal conductivity $\sigma_{xx}$.
Across all stacks and Fermi energies, $\sigma_{xx}$ remains finite with no tendency to vanish in the regime where $\sigma_{xy}$ is already quantized.
In addition, $\sigma_{xx}$ increases with the cone order $N$, consistent with the larger low-energy density of states of the chiral Lifshitz fermion.
The finite longitudinal conductivity is part of the gapless response, since the Hall signal originates from massless fermions on the parity-invariant Fermi surface, which necessarily also support longitudinal transport \cite{Zou2022PRB,Zou2023PRB,Shen2026PRB}.
This is in contrast to a quantum anomalous Hall insulator, where the quantized Hall plateau is accompanied by a vanishing $\sigma_{xx}$ \cite{Nagaosa2010RMP,ChangLiuMacDonald2023RMP}; the distinction persists even for the AB and ABCA stacks, where $W/2$ takes the integer values $1$ and $2$.

In summary, dephasing-assisted six-terminal transport provides a transport check of the parity-anomalous Hall response of chiral Lifshitz fermions.
Under strong dephasing, the AB, ABC, and ABCA stacks approach Hall plateaus fixed by the Fermi-loop winding, $\sigma_{xy}=W/2$ in units of $e^2/h$, while $\sigma_{xx}$ remains finite.
The coexistence of a quantized Hall response with finite longitudinal conductivity identifies a gapless semimetallic Hall state \cite{8vs2-jvc4, Liu2026PRB}, rather than an ordinary quantum anomalous Hall insulator whose plateau is accompanied by vanishing $\sigma_{xx}$.

\section{Summary and Discussion}
\label{sec:summary-discussion}

In summary, we studied the parity-anomalous Hall response associated with
chiral Lifshitz fermions. The
low-energy object is a parity-invariant Fermi loop whose in-plane pseudospin
winds \(W=N\) times, producing a Berry phase \(\pi W\). In a
winding-preserving lattice completion, this target winding must be accompanied
by compensating high-energy zeros. Removing these compensating modes from the
low-energy spectrum requires high-energy parity-breaking regularization masses, while the target
chiral Lifshitz fermion remains parity invariant. The resulting anomalous response is $\sigma_{xy}^{\mathrm{anom}}=W e^2/2h$. In this sense, the chiral Lifshitz fermion acts as $N$ same-chirality Dirac cones merged at one momentum point, whose Fermi-loop contributions add constructively.

Multilayer Haldane models provide a concrete realization of this theory. Tuned to the valley-selective critical line, the AB, ABC, and ABCA stacks host a single gapless chiral Lifshitz fermion at the $K$ valley with quadratic, cubic, and quartic dispersion, while the gapped $K'$ valley and the high-energy bands supply the parity-breaking regularization. The projected pseudospin textures give $W=2$, $3$, and $4$. The Kubo Hall conductance forms plateaus at $\sigma_{xy}=1$, $3/2$, and $2$ in units of $e^2/h$; the equilibrium boundary current converges to the same $W/2$; and dephasing-assisted six-terminal transport reproduces the quantized plateaus in a Hall-bar geometry.

The resulting state is a gapless, semimetallic anomalous Hall state rather
than a Chern insulator. Its Hall response is carried by the massless chiral
Lifshitz Fermi loop and therefore coexists with finite longitudinal
conductivity. This distinguishes it from a quantum anomalous Hall background
plus a single massless Dirac cone: the latter can have the same total Hall
coefficient, but its integer part is carried by filled Chern bands and chiral
edge modes. Here, by contrast, the full \(W/2\) contribution is tied to the
winding of the gapless Fermi loop. This distinction persists even for the AB
and ABCA cases, where \(W/2\) takes the integer values \(1\) and \(2\). Our
results support the interpretation that, in the multilayer Haldane realization,
the Hall coefficient is fixed by the winding of a gapless parity-invariant
chiral Lifshitz Fermi loop, rather than by a bulk insulating Chern number.

{\color{black}
\hypersetup{linkcolor=blue,citecolor=blue}
In our lattice realization, the massless chiral Lifshitz node is obtained
by tuning the valley mass to zero. Electron--electron interactions may
shift this critical condition or favor ordered phases that reconstruct
the node. Interaction-driven gapped and nematic phases in
quadratic-band-crossing models illustrate the latter
possibility~\cite{SunYaoFradkinKivelson2009}. An important next step is to
determine whether the gapless state remains accessible by tuning in an
interacting system, and how its Hall response is modified. These questions
require a specified microscopic interaction model.

The Fermi-loop criterion also guides materials selection: a nonzero pseudospin or
orbital winding is essential, whereas nonlinear dispersion alone is
insufficient. Relevant starting points include chiral multilayer
graphene~\cite{MinMacDonald2008,KoshinoMcCann2009}, orbital quadratic
crossings in (111) oxide
heterostructures~\cite{RueggFiete2011Orbital,Ruegg2013Distortion}, and
kagome-derived quadratic
crossings~\cite{SunYaoFradkinKivelson2009,Kang2020CoSn}.
These systems illustrate how nonlinear dispersion accompanied by nonzero
pseudospin or orbital winding can arise in condensed matter. In CoSn,
spin--orbit coupling opens a gap at the quadratic
touching~\cite{Kang2020CoSn}. Accessing the proposed Hall state calls for
tuning a single chiral sector to a massless, parity-invariant Fermi loop,
while the remaining bands provide the parity-breaking lattice
regularization.
\par}

\section*{Acknowledgments}
This work was supported by National Natural Science Foundation of China (Grant Nos. 12474057 and 124B2069) and the Fundamental Research Funds for the Central Universities. The computational resources were supported by the High-Performance Computing Platform of
Beihang University and Peking University.

\appendix
{\color{black}
\section{Smr\v{c}ka--St\v{r}eda decomposition of the intrinsic Hall conductivity}
\label{app:streda}
\begingroup
\color{black}
\hypersetup{linkcolor=blue,citecolor=blue}

This appendix relates the intrinsic Kubo response in
Eqs.~\eqref{eq:kubo-berry} and \eqref{eq:kubo} to the conventional
Smr\v{c}ka--St\v{r}eda decomposition,
$\sigma_{xy}=\sigma_{xy}^{I}+\sigma_{xy}^{II}$~\cite{Bonbien2020PRB}.
We derive both contributions in the clean limit and show that their
sum reproduces the occupied-state Berry-curvature expression used
in the main text.

\begin{figure}[thbp]
\centering
\includegraphics[width=\linewidth]{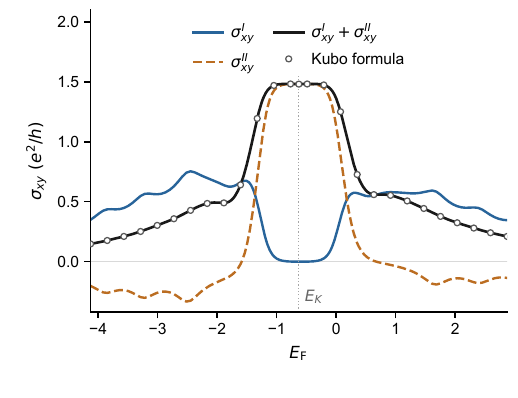}
\caption{\color{black}Smr\v{c}ka--St\v{r}eda decomposition of the intrinsic bulk
Hall conductivity for the ABC stack. The blue solid and orange dashed
curves show $\sigma_{xy}^{I}$ and $\sigma_{xy}^{II}$, respectively;
the black solid curve is their sum. Open circles show the direct
evaluation of Eq.~\eqref{eq:kubo}.
The vertical dotted line marks the
gapless energy $E_K$.}
\label{fig:streda}
\end{figure}

We use the dimensionless reciprocal coordinates
$q_a=\mathbf{k}\cdot\mathbf{a}_a$ and the Hamiltonian derivatives
$V_a=\partial H_N/\partial q_a$, with $a=1,2$.
The same periodic-cell Bloch Hamiltonian and matrix derivatives
are used for the two terms and for the direct evaluation of
Eq.~\eqref{eq:kubo}. We define
\[
\langle X\rangle_q=\int_0^{2\pi}\!\int_0^{2\pi}
\frac{dq_1\,dq_2}{(2\pi)^2}\,X,
\qquad s_{xy}^{\nu}=\frac{\sigma_{xy}^{\nu}}{e^2/h},
\]
with $\nu=I,II$.
For the positively oriented primitive vectors used here, the
coordinate Jacobians cancel between the integration measure
and the antisymmetric product of the two derivative vertices.
No additional unit-cell-area factor is required below.
The antisymmetric Hall parts of the two terms can be written as
\begin{align}
s_{xy}^{I}
&=\left\langle\int d\epsilon\,f'(\epsilon)
\mathcal{A}_{12}\operatorname{Re}\operatorname{Tr}
\left[V_1G^RV_2(G^R-G^A)\right]\right\rangle_q,
\label{eq:streda-I-trace}\\
s_{xy}^{II}
&=\left\langle\int d\epsilon\,f(\epsilon)
\mathcal{A}_{12}\operatorname{Re}\operatorname{Tr}
\left[V_1G^RV_2\partial_\epsilon G^R\right.
\right.\nonumber\\[-1ex]
&\hspace{37mm}\left.\left.
-V_1\partial_\epsilon G^RV_2G^R\right]\right\rangle_q.
\label{eq:streda-II-trace}
\end{align}
Here $G^{R/A}(\epsilon,\mathbf q)
=[\epsilon-H_N(\mathbf q)\pm i\eta_G]^{-1}$, and the expressions
are understood in the clean spectral limit $\eta_G\to0^+$.
The trace runs over all layer and sublattice orbitals, and
$\mathcal{A}_{12}$ takes half the difference between an expression
and the one with $V_1$ and $V_2$ interchanged. We use
$f(\epsilon)=\{1+\exp[(\epsilon-\EF)/(k_BT)]\}^{-1}$ and
$f'(\epsilon)=\partial_\epsilon f(\epsilon)$.
The infinitesimal regulator $\eta_G$ is distinct from the finite
strip spectral width $\eta$ in Eq.~\eqref{eq:equilibrium-current}.

At a nondegenerate momentum point, define
$\Delta_{nm}=E_n-E_m$ and
$P_{nm}=\operatorname{Im}[(V_1)_{nm}(V_2)_{mn}]$ in the band basis.
Taking $\eta_G\to0^+$ at fixed $f$ and performing the energy
integrations gives
\begin{align}
s_{xy}^{I}
&=2\pi\left\langle\sum_{n<m}
P_{nm}\frac{f'_n+f'_m}{\Delta_{nm}}\right\rangle_q,
\label{eq:streda-I-bands}\\
s_{xy}^{II}
&=4\pi\left\langle\sum_{n<m}P_{nm}
\left[\frac{f_n-f_m}{\Delta_{nm}^{2}}
-\frac{f'_n+f'_m}{2\Delta_{nm}}\right]\right\rangle_q,
\label{eq:streda-II-bands}
\end{align}
where $f_n=f(E_n)$ and $f'_n=f'(E_n)$. The sums include all
distinct band pairs. The terms proportional to $f'_n+f'_m$
cancel in the sum, yielding
\begin{equation}
s_{xy}^{I}+s_{xy}^{II}
=4\pi\left\langle\sum_{n<m}
P_{nm}\frac{f_n-f_m}{\Delta_{nm}^{2}}\right\rangle_q,
\label{eq:streda-sum}
\end{equation}
which is the occupied-state Berry-curvature expression in
Eqs.~\eqref{eq:kubo-berry} and \eqref{eq:kubo}.
Equation~\eqref{eq:streda-sum} therefore includes both
Smr\v{c}ka--St\v{r}eda contributions.
The labels I and II refer to this Green-function decomposition
and should not be identified with the separate filled-band and
Fermi-loop contributions discussed in
Sec.~\ref{subsec:n-half-derivation}~\cite{Bonbien2020PRB}.

Figure~\ref{fig:streda} illustrates the decomposition for the ABC stack
with the same lattice parameters as in Sec.~\ref{subsec:hall-current}.
We evaluate both band-sum expressions on an $800\times800$ shifted
momentum mesh that avoids the band-touching points.
The open circles show the reference calculation on a
$501\times501$ mesh. The sum
$\sigma_{xy}^{I}+\sigma_{xy}^{II}$ agrees with these reference
data, with a maximum absolute difference of
$4.0\times10^{-4}e^2/h$ at the reference energies.
Although $\sigma_{xy}^{I}$ is small
near the plateau center, it is finite away from that region
and is retained throughout the calculation.

We also checked momentum-mesh convergence using shifted meshes of
$250\times250$, $500\times500$, and $800\times800$. At nine representative
Fermi energies, the maximum change in the total Hall conductivity
between the two finest meshes is approximately $2.3\times10^{-4}e^2/h$.

\endgroup
}

\bibliography{main}

\end{document}